\documentclass[conference]{IEEEtran}
\IEEEoverridecommandlockouts

\usepackage{cite}
\usepackage{amsmath}
\usepackage{amssymb}
\usepackage{amsfonts}
\usepackage{graphicx}
\usepackage{multirow}
\usepackage{textcomp}
\usepackage{algorithm}
\usepackage{algpseudocode}
\usepackage[table]{xcolor}
\def\BibTeX{{\rm B\kern-.05em{\sc i\kern-.025em b}\kern-.08em
    T\kern-.1667em\lower.7ex\hbox{E}\kern-.125emX}}
\begin{document}

\title{Formal Verification of Parameterized Systems based on Induction\\
}

\author{\IEEEauthorblockN{1\textsuperscript{st} Jiaqi Xiu}
\IEEEauthorblockA{\textit{Key Laboratory of System Software} \\
\textit{Institute of Software, Chinese Academy of Sciences}\\
Beijing, China \\
xiujq@ios.ac.cn} \\
\and
\IEEEauthorblockN{2\textsuperscript{nd} Yongjian Li\dag}
\IEEEauthorblockA{\textit{Key Laboratory of System Software} \\
\textit{Institute of Software, Chinese Academy of Sciences}\\
Beijing, China \\
lyj238@ios.ac.cn}
}


\maketitle

\begin{abstract}
Parameterized systems play a crucial role in the computer field, and their security is of great significance. Formal verification of parameterized protocols is especially challenging due to its ``parameterized" feature, which brings complexity and undecidability. Existing automated parameterized verification methods have limitations, such as facing difficulties in automatically deriving parameterized invariants constrained by mixed Forall and Exists quantifiers, or having challenges in completing the parameterized verification of large and complex protocols.
This paper proposes a formal verification framework for parameterized systems based on induction, named wiseParaverifier. It starts from small concretizations of protocols, analyzes inductive counterexamples, and constructs counterexample formulas to guide the entire process of parameterized verification. It also presents a heuristic Generalize method to quickly find auxiliary invariants, a method for promoting complex mixed quantifiers and merging parameterized invariants, and uses symmetric reduction ideas to accelerate the verification process.
Experimental results show that wiseParaverifier can successfully complete automatic inductive verification on 7 cache coherence protocols and 10 distributed protocols. It has strong verification capabilities and migration capabilities, and can provide concise and readable verification results, which is helpful for learners to understand protocol behaviors.
\end{abstract}


\section{Introduction}
Parameterized systems play a crucial role in the computer field, modeling parameterized systems leads to the formation of parameterized protocols. The parameters within these systems typically represent system configurations or environmental conditions, such as variables that can affect system behavior, including the number of processes, resource limitations, the number of users, or network nodes. Common types of parameterized protocols include cache coherence protocols, distributed protocols, security protocols, and network communication protocols, which are widely applied in areas such as hardware design verification and multi-hreaded programs \cite{abdulla2007parameterized,abdulla2016parameterized}.

In modern computer architectures, cache coherence protocols are key technologies for ensuring the consistency of cached data in multi-core or multi-processor systems. With the development of computer hardware technology, the increasing number of processors has made the parallelism of caches an important means of enhancing system performance \cite{fujiki2019duality,mallayev2021cache,eshel2010panache,kunjir2017robus}. Distributed protocols are core technologies for ensuring correct and efficient data exchange and coordination among multiple computing nodes. With the rapid development of technologies such as cloud computing and big data, the application scenarios of distributed systems have become more extensive, playing important roles in file systems, high-performance computing, and database systems \cite{ghemawat2003google,chang2008bigtable,dean2008mapreduce}.

In these parameterized systems, security is a primary consideration, especially in systems involving sensitive data processing, network communication, and multi-user environments. Therefore, the formal verification of parameterized protocols, namely parameterized verification, has attracted extensive attention from both academia and industry \cite{apt1986limits}. The key in verifying parameterized protocols lies in finding a set of auxiliary invariants that, together with the security properties, can form inductive invariants, thereby completing the parameterized verification.

The challenge in the inductive proof of parameterized systems is that although it is relatively straightforward to identify inductive invariants for certain small scale concretizations of the system, it does not imply that these invariants will retain their inductive characteristics when applied to larger scale system concretizations. To address this issue, researchers have proposed many methods to automate this process. The first type, inductive verification starting from small concretizations \cite{lv2007computing,li2016novel,goel2021towards,hance2021finding,goel2021symmetry}, but may face the state space explosion problem, has high requirements for model checkers, and struggles to generalize complex invariants withmixed Forall and Exists quantifiers. The second type, using data-driven techniques to directly construct parameterized auxiliary invariants \cite{yao2021distai,yao2022duoai,xia2025discovering}, has a large enumeration volume, is prone to explosion of enumeration in complex protocol verification, and the auxiliary invariants found are less concise and less readable. Futhermore, most current automated verification tools can only handle single type parameterized protocols.

Limited by the scale and the diversity of protocols, developing a general and efficient automated parameterized verification system has emerged as an urgent issue that demands immediate resolution. In response to this problem, this paper conducts idepth research. Starting from small examples, we proposed a counter example guided inductive verification method. The main contributions of this paper are as follows:
\begin{enumerate}
    \item A formal verification framework wiseParaverifier for parameterized systems based on induction is proposed. It begins with small concretizations of protocols, deeply analyzes inductive counter examples, and presents a method for constructing counter example formulas. 
    \item A heuristic Generalize method for quickly finding auxiliary invariants is proposed. The proposed method constructs heuristic information based on variable analysis of pre-invariants, enabling rapid identification of the candidate invariants set where the target auxiliary invariant lies. By combining incremental and decremental strategies, redundant terms are further eliminated to obtain the most concise description of the auxiliary invariant. 
    \item A method for promoting complex mixed quantifiers guided by type saturation and a theory for merging parameterized invariants are proposed. Aiming at the difficulty of accurately deriving Forall/Exists quantifier mixed constraints, this proposed method achieves fast and accurate inference of complex descriptions of Forall/Exists mixed quantifiers in parameterized invariants through fine grained grouping and expansion. 
\end{enumerate}

\section{Related Work}

\subsection{Verification Starting from Small Concretizations}

Researchers observed that inductive invariants often remain relatively stable as protocol parameters increase. Lv et al. \cite{lv2007computing} combined implicit invariants with the CMP method to use small protocol concretizations as reference instances for computing auxiliary invariants. L-CMP \cite{cao2018cmp} designed an automated parameterized verification framework based on association rule learning, which automatically learns auxiliary invariants from small concretizations of parameterized systems, abstracts them, and performs verification. In the verification of distributed protocols, I4 \cite{emerson2003exact} was the first work to automatically construct inductive invariants and complete parameterized verification. IC3PO \cite{goel2021towards,goel2021symmetry} also starts with invariants from small protocol concretizations and generalizes the inductive verification relationships of small concretizations to parameterized systems. SWISS \cite{hance2021finding} iteratively strengthens inductive invariants by adding invariants constructed from small concretizations until they are sufficient to prove parameterized protocols. 

These methods starting from small concretizations have achieved success in verifying cache coherence and distributed protocols. However, they struggle to verify both types of protocols effectively.

\subsection{Verification via Direct Construction of Parameterized Auxiliary Invariants}

Unlike methods starting from small concretizations, these approaches directly construct parameterized auxiliary invariants at the parameterized level. LIDO \cite{xia2025discovering} combines runtime monitoring and learning methods to generate effective invariant expressions through dynamic program analysis. Yao et al. \cite{yao2021distai,yao2022duoai} used data-driven methods and Ivy \cite{hawblitzel2015ironfleet} for auxiliary verification to directly construct and filter parameterized invariants. This approach is fast for verifying simple distributed protocols and can infer complex invariants with mixed Forall/Exists quantifiers based on invariant templates, automating the verification of several Paxos variants \cite{yao2022duoai}.Ivy is a successful automatic reasoning framework for inductive properties in recent years. However, since it is unable to automatically derive the inductive invariants of protocols, subsequent automation work only uses it as an auxiliary verification tool in the inductive relation proof stage.

These methods have an advantage in quantifier generalization since they avoid parameterizing invariants from concretizations to parameterized forms. However, their verification speed is often limited by heuristic information such as invariant templates.

\section{Method}

\begin{figure*}[!htbp]
    \centering
    \includegraphics[width=0.8\textwidth]{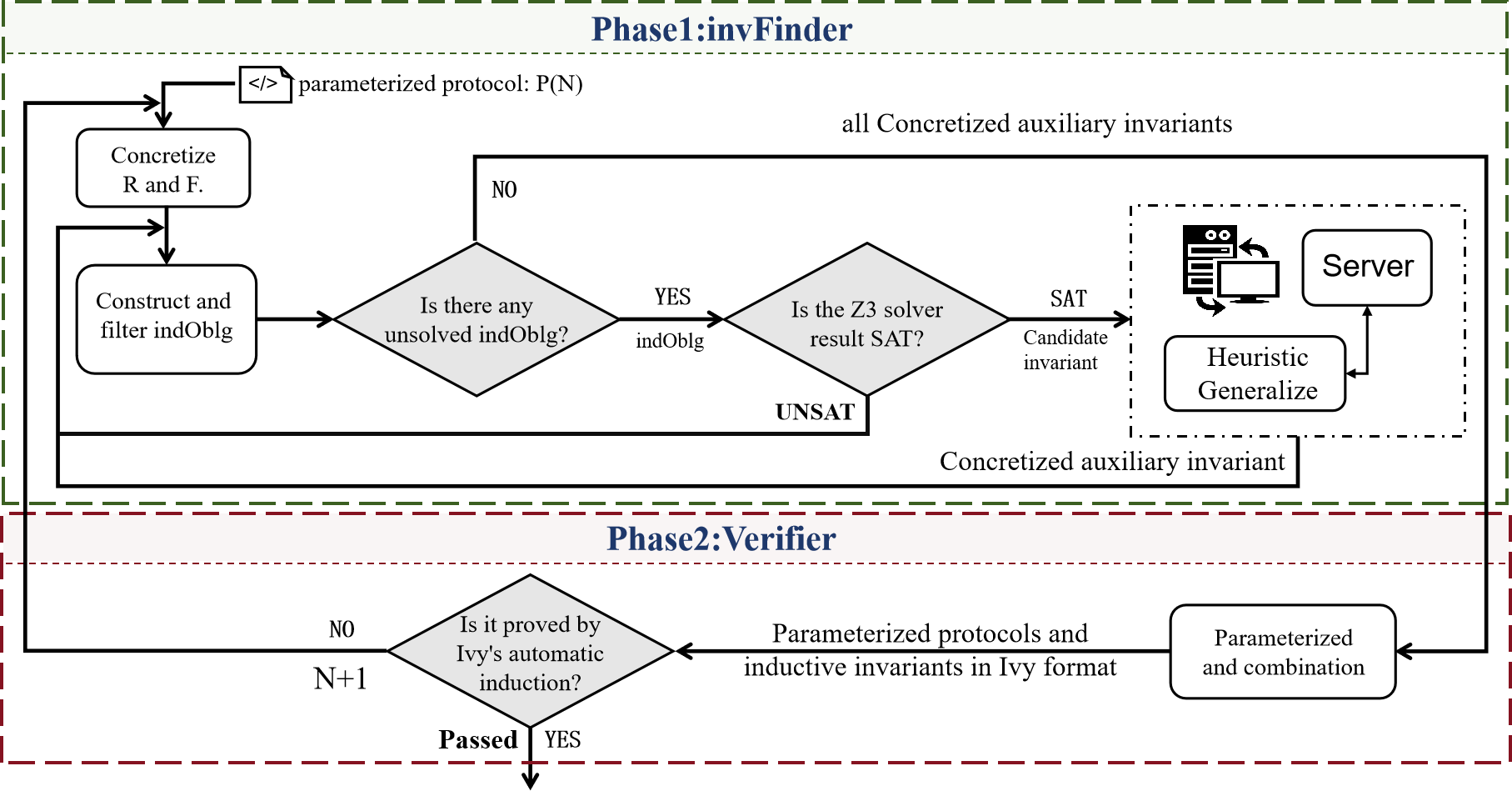}
    \caption{Workflow of wiseParaverifier}
    \label{fig:wise_flowchart}
\end{figure*}

As shown in Fig.~\ref{fig:wise_flowchart}, wiseParaverifier is an inductive formal verification method for parameterized systems that can automatically and efficiently search for and construct auxiliary invariants for the target safety properties of parameterized protocols, and complete inductive proofs of the protocols. The verification process consists of two stages: analyzing and constructing specific auxiliary invariants (invFinder), and generalizing and proving parameterized inductive invariants (Verifier).

\subsection{invFinder}
In this stage, the system will automatically analyze inductive counterexamples and construct specific auxiliary invariants for the input protocols. The auxiliary invariant determination is designed as a Client-Server architecture, where the server side encapsulates Murphi\cite{dill1996mur}, NuSMV and NuSMV-BMC\cite{cimatti2002nusmv} services. This stage has the following steps:

\subsubsection{Obtaining Minimum Concretizations of Transition Rules and Safety Properties}
Since wiseParaverifier uses inductive verification starting from small concretizations, the selection of initial sizes for protocols directly affects the efficiency of all inductive verification processes. The principles for selecting protocol concretization sizes are as follows:
\begin{enumerate}
    \item Concretization begins with the parameters in the invariant. First determine the minimum concretization based on the parameters required by the invariant, then determine the minimum concretization of the rules based on the parameter relationships between invariants and rules.
    \item The concrete protocol must be sufficient to simulate the same or symmetric transition behaviors as the protocol under any parameters.
    \item Among concretizations satisfying the above conditions, select the smallest concretization N.
\end{enumerate}

Assuming both invariant $F$ and rule $R$ contain only $type_1$ variables, we denotes the actual parameter count required by $F$ as $m$, and the actual parameter count required by $R$ as $k$. For simplicity, $type_1$ in $F$ can be instantiated as $1, 2, \cdots, m$, and $type_1$ in $R$ can be instantiated according to permutations of $1, 2, \cdots, n$. Although theoretically $n$ can take multiple values, to simplify the verification process and avoid duplicate concretizations, this section only considers the minimum $n$ satisfying the above relationships. Examples of instantiation for different $m$, $k$, and $n$ values include:
\begin{enumerate}
    \item If $m=2$ and $k=1$, $type_1$ in F can be concretized as $[1,2]$. For $R$, possible $type_1$ concretizations are $[1], [2], [3]$. The first two represent cases where $F$ and $R$ share the same concretization, while $[3]$ represents all cases where $F$ and $R$ have different concretization. The minimum n selected in this example would be 3.
    \item If $m=2 $and $k=2$, $type_1$ in $F$ is still concretized as $[1,2]$. For $R$, possible $type_1$ concretizations include $[1,2], [1,3], [2,1], [2,3], [3,1], [3,2], [3,4]$. Here $n=4$, but not all combinations are used due to symmetries or redundant meanings. For example, $[2,3]$ and $[2,4]$ both represent cases where the first parameter in $R$ matches the second parameter in $F$, while the second parameter in $R$ does not appear in $F$.
\end{enumerate}

\subsubsection{Constructing, Filtering, and Solving Counterexample Formulas}
To satisfy the inductive property of invariants, for any transition rule $R$ leading from state $s$ to $s'$, the invariant must hold in both $s$ and $s'$. If not, the inductive continuity of the invariant is violated. From a counterexample perspective, the counterexample formula indOblg is constructed as follows:
\begin{equation}
    \begin{aligned}
        \mathrm{indOblg (R,F)}&\equiv\mathrm{guard} (R) \wedge \\
        &\mathrm{asgn2Form (act} ( R') )\wedge \\
        &\mathrm{assumptions}( R,F)\wedge\neg F'
    \end{aligned}
\end{equation}

Where:
\begin{itemize}
    \item $\mathrm{guard}(R)$ represents the values of all relevant variables in the guard part of transition rule $R$ at state $s$.
    \item $\mathrm{asgn2Form( act}( R') )$ represents the assignments to all relevant variables in the action part a of transition rule $R'$ at state $s'$.
    \item $\mathrm{assumptions}( R,F) $ represents variables appearing in invariant $F$ but not in the action part a of transition rule $R$. All variables $var'$ in $\mathrm{assumptions}( R,F)$  are assigned values equal to $var$, implying variables unchanged by rule execution do not directly affect state $s'$.
    \item $\neg F'$ indicates the invariant will not hold at state $s'$, i.e., the system reaches a counterexample.
\end{itemize}

According to Hoare logic, if $s\vDash F$, then $s\vDash\mathrm{preCond}(F,a)$ when $s'$ is reachable via rule $R$. If $a$ does not modify any variables in $F$ ($s\leftrightarrow\mathrm{preCond}(F,a)$), then $F'$ trivially holds after rule execution, meaning no inductive counterexample can occur. wiseParaverifier automatically filters out such invalid indOblg formulas.

For each indOblg formula, wiseParaverifier creates a Z3 solver instance and adds constraints. Solvability is determined as follows:

\begin{enumerate}
    \item If unsolvable, $\mathrm{indOblg}( R,F)$ has no solution, meaning $\langle R, F\rangle$ does not form a CTI.
    \item If solvable, $\mathrm{indOblg}( R,F) $ has solutions consisting of variable value equations $eq$. In this case, $( R, F)$ forms a CTI. The solver returns the first valid solution, which wiseParaverifier uses to construct candidate invariants by negating the conjunction of $eq$ in $s$. The constructed auxiliary invariants follow the pattern in Eq.~\eqref{eq:invariant}, where $N$ is the protocol concretization and $W$ is the number of clause equations.
    \begin{equation}
        \label{eq:invariant}
        \Omega^N = \lnot \bigwedge_k^W f_k,~f_k = \text{eq}_i \lor \lnot \text{eq}_i
    \end{equation}
\end{enumerate}

\subsubsection{Heuristic Search Method for Auxiliary Invariants}
While indOblg formulas provide insights into inductive counterexamples, candidate invariants directly constructed from their solutions are often too complex. wiseParaverifier proposes an efficient heuristic Generalize strategy to filter irrelevant variables and construct concise auxiliary invariants.

\begin{figure}[H]
    \centering
    \includegraphics[width=0.3\textwidth]{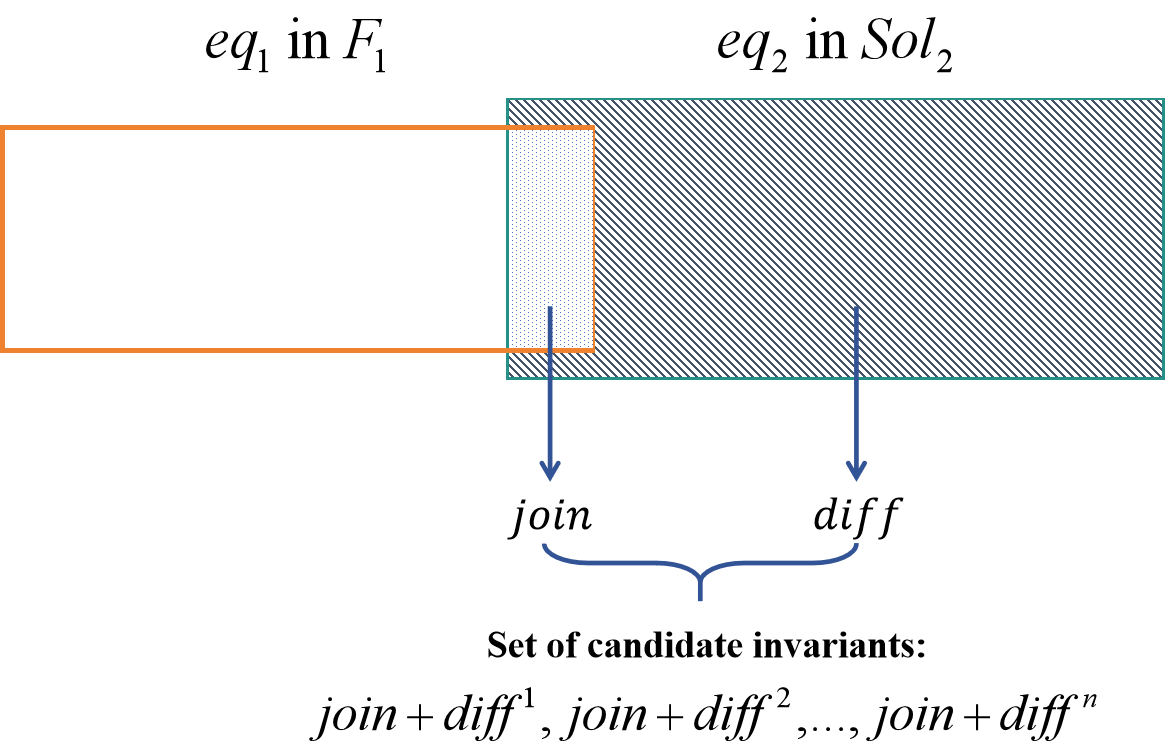}
    \caption{Illustration of the Core Idea of the Heuristic Generalize Algorithm}
    \label{fig:Heuristic}
\end{figure}

Empirical analysis indicates that new auxiliary invariants often relate to the $\neg F'$ terms in indOblg. As shown in Fig.~\ref{fig:Heuristic}, the heuristic Generalize algorithm uses variable information from previous invariants as guidance to rapidly identify relevant terms.

\begin{algorithm}
    \footnotesize
    \caption{\quad simplifyAuxInv}
    \label{alg:simplifyAuxInv}
    \begin{algorithmic}[1]
        \State \textbf{Inputs:} The indOblg set returned by the Z3 solver: $solution$.
        \State \textbf{Ouputs:} Whether $solution$ successfully constructs the auxiliary invariant. Returns its auxiliary invariant $auxInv$ if success.
        \State -- $simplifyAuxInvdecreasingly$: Iteratively remove irrelevant terms from the largest initial candidate.
        \State -- $simplifyAuxInvincreasingly$: Iteratively add valid terms to the smallest initial candidate.
        \Function{$simplifyAuxInvdecreasingly$}{$solution$}
            \State $canInv\gets turnSMVform(solution)$
            \State $passCheck\gets callNuSMV(canInv)$
            \If{$passCheck$}
                \State $auxInv\gets canInv$\
                \If{$solution.length > 1$}
                    \For{$subS \in getSublist(solution, solution.length - 1)$}
                        \If{$simplifyAuxInvdecreasingly[0]$}
                            \State \textbf{break}
                        \EndIf
                    \EndFor
                \EndIf
                \State \textbf{return} $True, auxInv$
            \EndIf
            \State \textbf{return} $False, None$
        \EndFunction

        \Function{$simplifyAuxInvincreasingly$}{$solution$}
            \For{$n \in solution.length$}
                \For{$subS \in getSublist(solution, n)$}
                    \State $auxInv\gets turnSMVform(solution)$
                    \State $passCheck\gets callNuSMV(auxInv)$
                    \If{$passCheck$}
                        \State \textbf{return} $True, auxInv$
                    \EndIf
                \EndFor
            \EndFor
            \State \textbf{return} $False, None$
        \EndFunction
    \end{algorithmic}
\end{algorithm}

Based on this, we proposes an efficient heuristic Generalize algorithm. Its core idea is to leverage variable information from pre-invariants as heuristic guidance to rapidly determine the variable terms of subsequent auxiliary invariants.

As shown in Fig.~\ref{fig:Heuristic}, $F_1$ represents pre-invariants known or previously derived by the wiseParaverifier system, while $Sol_2$ is the solution set of counterexamples for the candidate invariant being solved. The equation terms in $F_1$ and $Sol_2$ form sets $eq_1$ and $eq_2$ respectively. Typically, there exists a subset of identical equation terms join between $eq_1$ and $eq_2$. This $join$ set serves as heuristic guidance for constructing the minimal auxiliary invariant from $Sol_2$. Meanwhile, the unique equation terms $diff$ in $Sol_2$ are incrementally added to the $join$ set as subsets to form candidate invariant subsets. The notation $diff^n$ in Fig.~\ref{fig:Heuristic} denotes permutations of selecting $n$ equation terms from the $diff$ set. When the external model checker determines that a candidate invariant subset $join+diff^n$ contains the correct protocol invariant, it is considered the target auxiliary invariant derived from $Sol_2$, and the computation of $join+diff^{n+1}$ stops.

However, the above determination process does not encompass the full candidate invariants determination process involved in the heuristic Generalize algorithm. This is because although the invariant information set $join$ provides heuristic guidance during the Generalize process, not all equation terms in $join$ are effective. Nevertheless, the remaining candidates are already very close to the target auxiliary invariant. At this point, combining with the two auxiliary invariant search algorithms described in Algo.~\ref{alg:simplifyAuxInv} allows for the rapid elimination of remaining irrelevant terms.In Algo.~\ref{alg:simplifyAuxInv}, the functions simplifyAuxInvdecreasingly and simplifyAuxInvincreasingly represent the Decreasing strategy and the Increasing strategy respectively, and each of them has its own advantages in different scenarios. 

The process of the Generalize heuristic algorithm is shown in Algo.~\ref {alg:Generalization}. This algorithm can be divided into four stages:

\begin{enumerate}
    \item Heuristic search for the candidate invariants set containing the target auxiliary invariant (Lines 13-17 of Algo.~\ref{alg:Generalization}).
    \item Final determination of the target auxiliary invariant by combining with Algo.~\ref{alg:simplifyAuxInv} (Lines 18-21 of Algo.~\ref{alg:Generalization}).
    \item Adding constraints to accelerate counterexample solving and invariant search (Lines 24-25 of Algo.~\ref{alg:Generalization}). wiseParaverifier adds constraints to the Z3 solver after each auxiliary invariant is solved to speed up subsequent processes.
    \item A fallback non-heuristic Generalize method (Lines 30-32 and 3-11 of Algo.~\ref{alg:Generalization}) to ensure verification rigor for potential invariants not found by the heuristic approach, using the simplifyAuxInv algorithm described in Algo.~\ref{alg:simplifyAuxInv}
\end{enumerate}

\subsection{Verifier}
\subsection{Parameterization and Merging of Auxiliary Invariant Concretizations}
The auxiliary invariants obtained by invFinder need to be  parameterized in the Verifier stage to complete the parameterized verification. The difficulty in this stage lies in determining the correct quantifier constraints that should be added to each parameter type after generalizing the specific invariants. wiseParaverifier employs Ivy for automatic inductive verification, thus the results of parameterization in the described Verifier stage are all in Ivy format.
    
\begin{algorithm}[H]
    \footnotesize
    \caption{Generalization}
    \label{alg:Generalization}
    \begin{algorithmic}[1]
        \State \textbf{Inputs:} Z3 solver $SmtSolver$, and its returned indOblg solution set $solution$.
        \State \textbf{Ouputs:} target concise auxiliary invariant $auxInv$.
        \Function{$regular\_generalize$}{$solution, SmtSolver$}
            \State $auxInv\gets simplifyAuxInv(solution)$
            \If{$auxInv \neq None ~\land ~isLegal(auxInv)$}
                \State $addSMTAssertions(auxInv, SmtSolver)$
            \Else
                \State $auxInv\gets None$
            \EndIf
            \State \textbf{return} $auxInv$
        \EndFunction
        \Function{$heuristic\_generalize$}{$join, diff, solution, SmtSolver$}
            \If{$join \neq None$}
                \For{$n \in diff.length$}
                    \For{$subS \in getsublist(diff, n)$}
                        \State  $canInv\gets turnSMVform(subS \cup join)$
                        \State $passCheck\gets callNuSMV(canINv)$
                        \If{$passCheck$}
                            \State $auxInv\gets simplifyAuxInv(canInv)$
                            \State \textbf{break}
                        \EndIf
                    \EndFor
                \EndFor
                \If{$auxInv \neq None ~\land ~isLegal(auxInv)$}
                    \State $addSMTAssertions(auxInv, SmtSolver)$
                \Else
                    \State $auxInv\gets None$
                \EndIf
            \EndIf
            \If{$join ~is~ None~ \lor~ auxInv ~is~ None$}
            \State $auxInv\gets regular\_generalize(solution, SmtSolver)$
            \EndIf
            \State \textbf{return} $auxInv$
        \EndFunction
    \end{algorithmic}
\end{algorithm}

\textbf{Hybrid Quantifier Generalization Based on Type Saturation}\quad The proposed method developed a parameterized invariant quantifier deriving method based on parameter type saturation characteristics, which can rapidly and accurately infer complex Forall and Exists quantifier descriptions in parameterized invariants.

In this method, $C(t)=n_t$ is defined as the concrete size of parameter type $t$ in protocol $P(N)$, and $\#(\Omega,t)$ represents the number of constant concretizations of parameter type $t$ appearing in clause $\Omega$. Based on this, the type saturation $\gamma(t)$ of parameter type $t$ is defined as follows:
\begin{equation}
    \mathrm \gamma(t)= \frac{\#(\Omega,t)}{C(t)}
\end{equation}

The type saturation $\gamma(t)$ ranges from $(0,1]$. Based on whether $\gamma(t)=1$, the generalization cases can be divided into the following two scenarios:

\begin{enumerate}
    \item When $\gamma(t)\neq1$ (i.e., $\#(\Omega,t)<C(t)$), all concretizations of type $t$ in $P(N)$ do not fully participate in clause $\Omega$. Increasing $C(t)$ in $P(N)$ does not affect $\#(\Omega,t)$, meaning the occurrence of type $t$ in $\Omega$ is independent of its concrete size. Even concretizing type $t$ to larger values only produces symmetrically equivalent clauses to small concretizations.
    
    Parameter types with this saturation range will be described using Forall quantifiers in $\Omega$'s parameterization.
    \item When $\gamma(t)=1$ (i.e., $\#(\Omega,t)=C(t)$), all concretizations of type $t$ in $P(N)$ must participate in clause $\Omega$. The direct relationship between $\#(\Omega,t)$ and $C(t)$ cannot be determined immediately, as the correctness of the concrete invariant may depend on specific permutations of the current parameter concretize scale. Directly generalizing to larger scales ($N>C(s)$) may violate original constraints.Due to potential behavioral changes during concretization scaling, the correctness of clause $\Omega$ is reverified on $P(N+1)$, leading to two further generalization cases:
    \begin{enumerate}
        \item If $\Omega$ still holds, indicating symmetric behavior and invariants before/after scaling, type $t$ in the parameterized clause $\Omega$ uses Forall quantifiers as in Case 1.
        \item If $\Omega$ does not hold, the new concretization break the small concretization constraints, requiring Exists quantifiers to limit the scope. However, analysis of numerous protocol concretizations shows that even in this scenario, $t$ may be partially constrained by Forall quantifier, while the other part is constrained by the Exists quantifier. 
    \end{enumerate}
\end{enumerate}

Therefore, it is necessary to perform more fine-grained quantifier parameterization grouping for all specific  $t_1,t_2\cdots t_n$ of type $t$. Treating $t_1,t_2\cdots t_n$ as minimal groupings of type $t$, the extended invariant for each grouping g is constructed for the concretization $P(N+1)$ using the following formula:
\begin{equation}
    \label{eq:omegaG}
    \Omega_g'=\Omega\left[g^N \mapsto g^{N+1}\right],~\Omega'=\Omega_g'\land\Omega
\end{equation}

The extension principle of $g^N \mapsto g^{N+1}$ in Eq.~\eqref{eq:omegaG} is to supplement all possible parameter concretizations of the types involved in the group $g$ within $g^{N+1}$, as shown in Eq.~\eqref{eq:gN+1}.

\begin{equation}
    \label{eq:gN+1}
    g^N \mapsto g^{N+1} = \bigwedge_i^{N_t+1} g_i
\end{equation}

The pattern of the clause $\Omega$ before extension is expressed in Eq.~\eqref{eq:invariant}, while the extended clause $\Omega'$ follows the form in Eq.~\eqref{eq:Omega'}.

\begin{equation}
\label{eq:Omega'}
    \Omega' = \lnot \left ( \bigwedge_k^W f_k \land  \bigwedge_i^{N_t+1} g_i \right )
\end{equation}

Introducing the existential quantifier Exists to transform Eq.~\eqref{eq:Omega'} further yields an equivalent formula description expressed in Eq.~\eqref{eq:ExistsOmega'}.

\begin{equation}
\label{eq:ExistsOmega'}
    \Omega' =\exists i:{t_1,t_2,..t_{N_t+1}}. \lnot \left ( \bigwedge_k^W f_k \land   g_i \right )
\end{equation}

If $P(N+1)\nvDash \Omega'$, it indicates that the behavior of grouping $g$ can be preserved in the extended concretization, and the constraints derived from small concretizations can be applied to the extended one. The concretized $t_g$ represented by this grouping will be described using Forall quantifiers after parameterization. Conversely, if $P(N+1)\vDash \Omega'$, it means the grouping $g$ depends on specific concretized descriptions, and the constraints from small - scale concretizations cannot be directly applied to the extended concretization. The concretized $t_g$ represented by this grouping will be described using Exists quantifiers after parameterization. Notably, when multiple extended groupings of type $t$ satisfy $P(N+1)\vDash \Omega'$, the parameterized concretized $t_g$ of each grouping will be connected by $\lor$.

\textbf{Parameterized Invariant Merging}\quad The principle for merging and streamlining parameterized invariants is as follows: If the variable types and equation terms described by the Forall quantifier in the clauses are exactly the same, and the variable types described by the Exists quantifier are the same, then merging can be considered. Denote the merged parameterized invariant as $\phi$. wiseParaverifier can perform the merging of parameterized invariants in the following scenarios:

\begin{enumerate}
    \item Simple Equivalence Merging. Some different auxiliary invariant concretizations may parameterize to identical parameterized invariants via Hybrid Quantifier Generalization Based on Type Saturation. For such invariant groups, retaining one representative as $\phi$ suffices.
    \item Implication Merging. When a group of parameterized invariants satisfies the implication relationship defined in Eq.~\eqref{eq:merge1}, select $\phi=\varphi\_k$:
    \begin{equation}
        \label{eq:merge1}
        \varphi\_k\Rightarrow\bigwedge_{i=1}^n\varphi\_i
    \end{equation}

    \item Constraint Strengthening Merging. When invariant combinations share identical term structures and quantifier patterns, a new stronger single constraint $\phi$ can replace the original multiple constraints $\bigwedge_{i=1}^n \phi_i$. The strengthened invariant $\phi$ and the multiple constraints $\bigwedge_{i=1}^n \phi_i$ must satisfy the relationship in Eq.~\eqref{eq:merge2} with the original invariants:
    \begin{equation}
        \label{eq:merge2}
         \phi \Rightarrow \bigwedge_{i=1}^n \varphi_i \quad and\quad \bigwedge_{i=1}^n \varphi_i \nRightarrow \phi
    \end{equation}
\end{enumerate}

\subsubsection{Symmetry Reduction Principles}
Although the heuristic Generalize algorithm saves significant computational overhead for constructing auxiliary invariants , including solver solving and external model checking tool calls, protocols with high complexity (e.g., Flash protocol) may still trap wiseParaverifier in state space explosion.

To address this challenge, we apply symmetry reduction principles from model checking to the auxiliary invariants solving phase of wiseParaverifier. Parameterized protocols contain abundant symmetric system structures and states, leading to symmetric relationships in auxiliary invariants derived from protocol CTIs. These symmetries introduce redundant computational costs during verification.

Algo.~\ref{alg:getSymmetryInvs} demonstrates how to construct all symmetric invariants for an auxiliary invariant. After generalization, the concretetized types may fall into three categories:
\begin{enumerate}
    \item Pure Forall-quantified types: Allow free permutations.
    \item Pure Existential-quantified types: Permutations are restricted to maintain correctness.
    \item Hybrid types: Combine Forall and Existential quantifiers.
\end{enumerate}
Algo.~\ref{alg:getSymmetryInvs} distinguishes these cases to efficiently generate symmetric invariants while preserving logical correctness.

\begin{algorithm}[!htbp]
    \footnotesize
    \caption{$getSymmetryInvs$: Symmetric Invariant Generation Algorithm}
    \label{alg:getSymmetryInvs}
    \begin{algorithmic}[1]
        \State \textbf{Inputs:} $oriInv$: auxiliary invariant get by Generalization
        \State \textbf{Ouputs:} $symmInvs$: All symmetric invariants of $oriInv$
        \State --Analyze variable/type concretization distributions and filter existential types
        \State $S \leftarrow Parse(oriInv)$
        \State $(D, V) \leftarrow AnalyzeVariables(S)$
        \State $ E \leftarrow FilterExistentialTypes(D,V)$
        \State --Generate permutations based on quantifier patterns
        \State $\_R = \times[Permute(V[t]) | t \notin E]$ \quad// Forall quantified types only
        \State $\_E\_P = \times[Permute(ExistVars(V[t])) | \in E]$ \quad// Existential parts of hybrid types
        \State $\_F\_P = \times[Permute(ForallVars(V[t])) | t \in E]$ \quad// Forall parts of hybrid types
        \State --Generate symmetric invariants through permutation combinations
        \For{$r \in R$}
            \For{$e \in E\_P$}
                \For{$f \in F_P$}
                    \State $S\_r = ApplyCombination(S, r)$
                    \State $S\_e = ApplyCombination(S\_r, e) \land S\_r$
                    \State $S\_final = ApplyForallPerm(S\_e, f) \land S_e$
                    \State $symmInvs \leftarrow SymmInvs \land Assemble(S\_final)$
                \EndFor
            \EndFor
        \EndFor
        \State \textbf{return} $symmInvs$
    \end{algorithmic}
\end{algorithm}

Since a single counterexample may generate multiple auxiliary invariants, many of which are equivalent or symmetric, the Generalize process risks becoming trapped in redundant invariant construction. To address this, symmetry reduction can be applied during Generalize to avoid generating equivalent or symmetric invariants.

\begin{algorithm}
    \footnotesize
    \caption{$addSMTAssertions$: SMT Acceleration Algorithm}
    \label{alg:addSMTAssertions}
    \begin{algorithmic}
        \State \textbf{Inputs:} $auxInv$: The auxiliary invariant  that have completed the Generalize process.
        \State \textbf{Inputs:} $SmtSolver$: The current instance of the Z3 solver.
        \State $symmInvs \leftarrow getSymmetryInvs(auxInv)$
        \For{$subI \in auxInv ~\cup~ symmInvs$}
            \State $SmtSolver.add(turnSMTAssertion(subI))$
        \EndFor
    \end{algorithmic}
\end{algorithm}

\begin{table*}[t]
    \caption{Automatic Parameterized Inductive Verification Results of wiseParaverifier}
    \centering
    \label{tab:comResults}
    \renewcommand{\arraystretch}{1.5}
    \resizebox{0.9\linewidth}{!}{
    \begin{tabular}{c|c c c c|c c c c}
         \hline
         \multirow{2}{*}{Protocols} & \multicolumn{4}{c|}{Automated Verification Completion Time (s)} & \multicolumn{4}{c}{Final Number of Parameterized Invariants} \\
         \cline{2-9}
         & \multicolumn{1}{c}{wiseParaverifier} & \multicolumn{1}{c}{Paraverifier} & \multicolumn{1}{c}{DuoAI} & \multicolumn{1}{c|}{IC3PO} & \multicolumn{1}{c}{wiseParaverifier} & \multicolumn{1}{c}{Paraverifier} & \multicolumn{1}{c}{DuoAI} & \multicolumn{1}{c}{IC3PO} \\
         \hline
         MutualExNodata & 1.46 & 1.98 & 1.01 & 1.82 & 4 & 5 & 14 & 5 \\
         MutualEx & 2.96 & 3.37 & 2.49 & 3.16 & 6 & 6 & 16 & 6  \\
         MutualEx\_M2 & 17.12  & error & 28.10 & 26.94 & 13 & error & 32 & 12  \\
         GermanNodata  & 19.14 & 21.76 & 535.80 & fail & 39 & 40 & 892 & fail\\
         German & 49.20 & 57.67 & fail & fail & 55 & 58 & fail & fail\\
         FlashNodata & 329.91 & 342.17 & fail & fail & 95 & 152 & fail & fail\\
         Flash & 585.68 & 589.23 & fail & fail & 135 & 162 & fail & fail\\
         \hline
         Decentralized\_lock & 4.66 & error & 9.38 &  5.98 & 15 & error & 16 & 4\\
         Lock\_server & 0.24 & error & 1.31 & 2.60 & 1 & error & 1 & 2\\
         Multi\_lock\_server & 2.55 & error & 1.94 & 11.05 & 11 & error & 12 & 9\\
         Ricart-Agrawala & 0.43 & error & 0.90 & 2.40 & 2 & error & 6 & 3\\
         Shard & 3.13 & error & 1.98 & 5.43 & 13 & error & 15 & 7\\
         Two\_phase\_commit & 4.23 & 6.18 & 2.12 & 5.29 & 10 & 10 & 7 & 11\\
         Client\_server\_ae(Exists)*  & 3.28 & error & 2.50 & 3.32 & 2 & error & 4 & 2\\
         Client\_server\_db\_ae(Exists)*  & 6.83 & error & 4.12 & fail & 8 & error & 7 & fail \\
         Toy\_consensus\_epr(Exists)* &  4.97 & error & 3.92 & 3.61 & 4 & error & 4 & 3 \\
         Consensus\_epr(Exists)* &  43.10 & error & 28.88 & 1248.26 & 16 & error & 7 & 8\\
         \hline
    \end{tabular}
    } \\
    \footnotesize{1. error: After modeling the protocol according to the method's requirements, the compilation process failed.} \\
    \footnotesize{2. fail: After inputting and modeling the protocol according to the method's requirements, the verification cannot be completed.} \\
    \footnotesize{3. (Exists)*: The protocol verification process involves parameterized types requiring the use of Forall/Exists hybrid quantifiers.}
\end{table*}

Specifically, after the Generalize algorithm constructs an auxiliary invariant $auxInv$, all its symmetric counterparts $symmInvs$ are precomputed. The combined constraint $auxInv \land symmInvs$ is then added to the Z3 solver as a new assertion to block future generation of symmetric invariants.

Finally, Algo.~\ref{alg:addSMTAssertions} incorporates all symmetric invariants and the original invariant into the Z3 solver during the Generalize process. This algorithm corresponds to the implementation of lines 6 and 25 in the Generalization procedure.

\section{Experiment}
We quantitative analysis the automated parameterized inductive verification capabilities of wiseParaverifier across 17 classic protocols, including 7 cache coherence protocols and 10 distributed protocols, with 4 distributed protocols requiring complex Forall/Exists hybrid quantifiers to describe parameterized auxiliary invariants. All experiments were conducted on an 18 cores Intel(R) Gold 6254 CPU @ 3.10GHz. Notably, the Server side typically only needed to run once, with relevant information of executed protocols recorded in its context.

\begin{table*}[tbh]
    \caption{Comparison Experimental Results of Different Generalize Algorithms and Their Usage Strategies}
    \centering
    \label{tab:compare_res}
    \renewcommand{\arraystretch}{1.5}
    \resizebox{0.92\linewidth}{!}{
    \begin{tabular}{c|cc|cc|cc|cc}
        \hline
        \multirow{3}{*}{Protocols} & \multicolumn{4}{c|}{Simple Generalize Algorithm} & \multicolumn{4}{c}{Heuristic Generalize Algorithm} \\
        \cline{2-9}
        & \multicolumn{2}{c|}{Increasingly Strategy Only} & \multicolumn{2}{c|}{Decreasingly Strategy Only} & \multicolumn{2}{c|}{Auxiliary Utilizing Increasingly Strategy} & \multicolumn{2}{c}{Auxiliary Utilizing Decreasingly Strategy} \\
        \cline{2-9}
        & Time (s) & call NuSMV & Time (s) & call NuSMV & Time (s) & call NuSMV & Time (s) & call NuSMV \\
        \hline
        MutualExNodata & \cellcolor{green!10}2.18 & 22 & 2.2 & 22 & \textbf{\textcolor{red}{1.46}} & 22 & \textbf{\textcolor{red}{1.46}} & 22 \\
        MutualEx & \cellcolor{green!10}4.09 & 55 & 4.21 & 44 & 3.89 & 46 & \textbf{\textcolor{red}{2.96}} & 40 \\
        MutualEx\_M2 & \cellcolor{green!10}83.77 & 227 & 85.02 & 227 & 20.9 & 266 & \textbf{\textcolor{red}{17.12}} & 226 \\
        GermanNoData & \cellcolor{green!10}32.45 & 477 & 39.45 & 356 & \textbf{\textcolor{red}{19.14}} & 297 & \textbf{\textcolor{red}{19.1}} & 272 \\
        German & \cellcolor{green!10}92.12 & 1319 & 97.27 & 1145 & 61.74 & 703 & \textbf{\textcolor{red}{49.20}} & 630 \\
        FlashNodata & / & / & / & / & 1669.69 & 1818 & \textbf{\textcolor{red}{329.91}} & 674 \\
        Flash & / & / & / & / & / & / & \textbf{\textcolor{red}{585.68}} & 965 \\
        \hline
        Decentralized\_lock & \cellcolor{green!10}7.03 & 120 & 7.05 & 109 & 5.03 & 75 & \textbf{\textcolor{red}{4.66}}& 60 \\
        Lock\_server & 0.25 & 4 & \cellcolor{green!10}\textbf{\textcolor{red}{0.24}} & 4 & 0.34 & 8 & 0.25 & 5 \\
        Multi\_lock\_server & \cellcolor{green!10}6.01 & 66 & 6.73 & 57 & 6.73 & 57 & \textbf{\textcolor{red}{2.55}} & 50 \\
        Ricart-Agrawala & 0.44 & 8 & \cellcolor{green!10}\textbf{\textcolor{red}{0.43}} & 8 & 0.54 & 10 & 0.54 & 10 \\
        Shard & \cellcolor{green!10}3.85 & 66 & 3.86 & 65 & 3.19 & 52 & \textbf{\textcolor{red}{3.13}} & 52 \\
        Two\_phase\_commit & \cellcolor{green!10}7.88 & 248 & 11.42 & 187 & 5.66 & 119 & \textbf{\textcolor{red}{4.23}} & 80 \\
        Client\_server\_ae(Exists)* & \cellcolor{green!10}8.46 & 159 & 9.31 & 89 & 9.31 & 72 & \textbf{\textcolor{red}{3.28}} & 45 \\
        Client\_server\_db\_ae(Exists)* & \cellcolor{green!10}11.68 & 210 & 13.58 & 175 & 11.33 & 154 & \textbf{\textcolor{red}{6.83}} & 125 \\
        Toy\_consensus\_epr(Exists)* & 12.71 & 121 & \cellcolor{green!10}12.08 & 97 & 6.04 & 89 & \textbf{\textcolor{red}{4.97}} & 88 \\
        Consensus\_epr(Exists)* & 183.23 & 2909 & \cellcolor{green!10}180.96 & 2093 & 72.48 & 869 & \textbf{\textcolor{red}{43.10}} & 753 \\
        \hline
    \end{tabular}
    } \\
    \footnotesize{/: The verification runtime has exceeded 24 hours, and the search for new counterexamples has not ceased.}
\end{table*}

\subsection{Comparative Study}
This work selects recent state-of-the-art approaches: Paraverifier, DuoAI, and IC3PO as benchmarks for comparing the verification capabilities of wiseParaverifier. Tab.~\ref{tab:comResults} presents a detailed comparison between wiseParaverifier and these advanced methods in terms of automated verification completion time and number of parameterized invariants required to achieve verification.

The qualitative results presented in Tab.~\ref{tab:comResults} demonstrate that the wiseParaverifier excels in the inductive verification of the automated protocol: 

\begin{enumerate}
    \item Paraverifier demonstrated strong performance in automated inductive verification of cache coherence protocols, successfully verifying 6 out of the 7 protocols in this work with completion times comparable to wiseParaverifier. Notably, it was the only state-of-the-art method in Tab.~\ref{tab:comResults} capable of verifying the complex Flash protocol automatically, albeit requiring significantly more auxiliary invariants than wiseParaverifier. However, Paraverifier failed to verify MutualEx\_M2 and all distributed protocols in the validation set due to its inability to handle high-dimensional array variables and infer Forall/Exists hybrid quantifiers.
    \item DuoAI, the first work to automate Paxos and several of its variants (excluding vertical Paxos), showed robust inductive verification capabilities. Table~\ref{tab:comResults} reveals that DuoAI achieved marginally faster average verification times than wiseParaverifier on distributed protocols. However, this difference was negligible due to the small state spaces and invariant candidate enumerations of these protocols. For cache coherence protocols with richer variable types and more rules, such as German and Flash, DuoAI not only incurred significantly longer runtimes and generated more invariants but also suffered from candidate invariant enumeration space explosion during verification.
    \item IC3PO performed well on distributed protocols but struggled with cache coherence protocols, verifying only 3 simple cases (the worst among the three baselines). It also exhibited limited ability to infer complex Forall/Exists hybrid quantifiers, completing verification for 3 out of 4 distributed protocols requiring such quantifiers. On the Consensus protocol, IC3PO’s runtime lagged significantly behind wiseParaverifier and DuoAI. Despite these limitations, IC3PO produced relatively concise verification results for most successfully verified protocols.
\end{enumerate}

In summary, wiseParaverifier demonstrates superior comprehensive verification capabilities, particularly in complex cache coherence protocols characterized by numerous rules, rich variable types, and large reachable state spaces, as well as distributed protocols involving high-dimensional array variables and intricate Forall/Exists hybrid quantifier inference.

\subsection{Ablation Study}
The designed Generalize algorithm is to construct the most concise auxiliary invariants from the solution set of the counterexample formula indOblg. To improve the efficiency of Generalize, we introduces the $join$ and $diff$ sets in the developed method, proposes a Generalize algorithm based on heuristic information, and uses the Decreasingly strategy and the Increasingly strategy as auxiliary algorithms for the heuristic Generalize algorithm.

To validate the effectiveness of the Heuristic Generalize Algorithm proposed in this section, Tab.~\ref{tab:compare_res} presents comparative experimental results for the 17 protocols in the validation set, detailing the automated verification completion time (in seconds) and number of external model checker invocations (NuSMV is used in the table) when wiseParaverifier employs different Generalize algorithms combined with auxiliary strategies during parameterized protocol verification.

As shown in Tab.~\ref{tab:compare_res}, red text highlights the shortest parameterized verification time for each protocol across all 4 Generalize algorithms and strategies, while light green text indicates the shortest time achieved using the simple Generalize algorithm with different strategy configurations. The ``/" symbol denotes protocols where verification runtime exceeded 86,400 seconds (24 hours) without halting counterexample discovery, deemed unable to complete automated parameterized verification under the corresponding algorithm strategy combination.

The comparative results in Tab.~\ref{tab:compare_res} reveal the following performance trends for the 17 protocols across 4 Generalize algorithms and strategies:

\begin{enumerate}
    \item The Heuristic Generalize Algorithm combined with the Decreasingly Strategy achieved the best performance, attaining the shortest verification time and least NuSMV invocations for 15 out of 17 protocols. Its verification times for the remaining two protocols (Lock\_server and Ricart-Agrawala) differed from the shortest by only 0.01s and 0.11s, respectively. Notably, this configuration was the only one capable of completing automated parameterized verification for all 17 protocols.
    \item The Heuristic Generalize Algorithm with the Increasingly Strategy performed well, achieving the shortest verification times for 2 protocols. It successfully verified 16 protocols, with the sole failure being the industrial-grade cache coherence protocol Flash, which features complex specifications, rich variable types, and an excessively large reachable state space. Verification times for other protocols (except FlashNodata) were comparable to the shortest recorded times.
    \item The simple Generalize Algorithm with the Increasingly Strategy achieved verification times close to the best for small protocols (MutualExNodata, Lock\_server), but incurred significantly longer runtimes for complex protocols (German, Consensus\_epr). This configuration verified 15 protocols but failed to handle large-scale protocols Flash and FlashNodata. Additionally, it required significantly more NuSMV invocations than the other three configurations.
    \item The simple Generalize Algorithm with the Decreasingly Strategy achieved the shortest times for 2 small, simple protocols (Lock\_server and Ricart-Agrawala), with negligible differences in runtime and NuSMV invocations compared to other strategies. It verified 15 protocols but failed to handle Flash and FlashNodata, similar to the previous simple algorithm configuration.
\end{enumerate}

The heuristic Generalize Algorithm efficiently constructs minimal auxiliary invariants from counterexample solutions, accelerating parameterized protocol verification in wiseParaverifier with fewer external model checker calls while handling industrial-grade protocols. The Decreasingly Strategy is optimal for heuristic generalization because its pre-pruned invariant set ($join \land diff^n$) typically requires only 1-2 redundant term removals. Light green highlights shortest verification times for simple Generalize Algorithm configurations where the Increasingly Strategy outperforms Decreasingly, though Decreasingly reduces NuSMV invocations, its complex invariants cause longer verification, whereas Increasingly's simpler, shorter invariants offset higher invocation counts with faster checks.

\section{Conclusion}

In this paper, an inductive formal verification method wiseParaverifier was proposed for parameterized systems that can automatically and efficiently search for and construct auxiliary invariants, and complete complete the inference of parameterized invariants and the proof of inductive relationships. By utilizing a heuristic Generalize method, wiseParaverifier could rapidly find auxiliary invariants. Furthermore, A method for promoting complex mixed quantifiers guided by type saturation and a theory for merging parameterized invariants are designed, mainly aiming at the difficulty of accurately inferring Forall/Exists quantifier mixed constraints. The proposed wiseParaverifier demonstrated robust automated parameterized inductive verification capabilities and strong transferability through multiple experimental groups, which include 7 cache coherence protocols and 10 distributed protocols.

\section*{Acknowledgment}

This research did not receive any specific grant from funding agencies in the public, commercial, or not-for-profit sectors.


\bibliographystyle{IEEEtran}
\bibliography{refs}

\begin{thebibliography}{10}
\providecommand{\url}[1]{#1}
\csname url@samestyle\endcsname
\providecommand{\newblock}{\relax}
\providecommand{\bibinfo}[2]{#2}
\providecommand{\BIBentrySTDinterwordspacing}{\spaceskip=0pt\relax}
\providecommand{\BIBentryALTinterwordstretchfactor}{4}
\providecommand{\BIBentryALTinterwordspacing}{\spaceskip=\fontdimen2\font plus
\BIBentryALTinterwordstretchfactor\fontdimen3\font minus \fontdimen4\font\relax}
\providecommand{\BIBforeignlanguage}[2]{{%
\expandafter\ifx\csname l@#1\endcsname\relax
\typeout{** WARNING: IEEEtran.bst: No hyphenation pattern has been}%
\typeout{** loaded for the language `#1'. Using the pattern for}%
\typeout{** the default language instead.}%
\else
\language=\csname l@#1\endcsname
\fi
#2}}
\providecommand{\BIBdecl}{\relax}
\BIBdecl

\bibitem{abdulla2007parameterized}
P.~A. Abdulla, G.~Delzanno, and A.~Rezine, ``Parameterized verification of infinite-state processes with global conditions,'' in \emph{Computer Aided Verification: 19th International Conference, CAV 2007, Berlin, Germany, July 3-7, 2007. Proceedings 19}.\hskip 1em plus 0.5em minus 0.4em\relax Springer, 2007, pp. 145--157.

\bibitem{abdulla2016parameterized}
P.~A. Abdulla and G.~Delzanno, ``Parameterized verification,'' \emph{International Journal on Software Tools for Technology Transfer}, vol.~18, pp. 469--473, 2016.

\bibitem{fujiki2019duality}
D.~Fujiki, S.~Mahlke, and R.~Das, ``Duality cache for data parallel acceleration,'' in \emph{Proceedings of the 46th International Symposium on Computer Architecture}, 2019, pp. 397--410.

\bibitem{mallayev2021cache}
O.~Mallayev, B.~Anvarjonov, and M.~Aziz, ``Cache problems in parallel computational processes,'' \emph{Annals of the Romanian Society for Cell Biology}, vol.~25, no.~3, pp. 8924--8934, 2021.

\bibitem{eshel2010panache}
M.~Eshel, R.~L. Haskin, D.~Hildebrand, M.~Naik, F.~B. Schmuck, and R.~Tewari, ``Panache: A parallel file system cache for global file access.'' in \emph{FAST}, vol.~10, 2010, pp. 1--14.

\bibitem{kunjir2017robus}
M.~Kunjir, B.~Fain, K.~Munagala, and S.~Babu, ``Robus: fair cache allocation for data-parallel workloads,'' in \emph{Proceedings of the 2017 ACM International Conference on Management of Data}, 2017, pp. 219--234.

\bibitem{ghemawat2003google}
S.~Ghemawat, H.~Gobioff, and S.-T. Leung, ``The google file system,'' in \emph{Proceedings of the nineteenth ACM symposium on Operating systems principles}, 2003, pp. 29--43.

\bibitem{chang2008bigtable}
F.~Chang, J.~Dean, S.~Ghemawat, W.~C. Hsieh, D.~A. Wallach, M.~Burrows, T.~Chandra, A.~Fikes, and R.~E. Gruber, ``Bigtable: A distributed storage system for structured data,'' \emph{ACM Transactions on Computer Systems (TOCS)}, vol.~26, no.~2, pp. 1--26, 2008.

\bibitem{dean2008mapreduce}
J.~Dean and S.~Ghemawat, ``Mapreduce: simplified data processing on large clusters,'' \emph{Communications of the ACM}, vol.~51, no.~1, pp. 107--113, 2008.

\bibitem{apt1986limits}
K.~R. Apt and D.~Kozen, ``Limits for automatic verification of finite-state concurrent systems,'' \emph{Inf. Process. Lett.}, vol.~22, no.~6, pp. 307--309, 1986.

\bibitem{lv2007computing}
Y.~Lv, H.~Lin, and H.~Pan, ``Computing invariants for parameter abstraction,'' in \emph{2007 5th IEEE/ACM International Conference on Formal Methods and Models for Codesign (MEMOCODE 2007)}.\hskip 1em plus 0.5em minus 0.4em\relax IEEE, 2007, pp. 29--38.

\bibitem{li2016novel}
Y.~Li, K.~Duan, Y.~Lv, J.~Pang, and S.~Cai, ``A novel approach to parameterized verification of cache coherence protocols,'' in \emph{2016 IEEE 34th International Conference on Computer Design (ICCD)}.\hskip 1em plus 0.5em minus 0.4em\relax IEEE, 2016, pp. 560--567.

\bibitem{goel2021towards}
A.~Goel and K.~Sakallah, ``Towards an automatic proof of lamport's paxos,'' in \emph{\# PLACEHOLDER\_PARENT\_METADATA\_VALUE\#}, 2021, pp. 112--122.

\bibitem{hance2021finding}
T.~Hance, M.~Heule, R.~Martins, and B.~Parno, ``Finding invariants of distributed systems: It's a small (enough) world after all,'' in \emph{18th USENIX symposium on networked systems design and implementation (NSDI 21)}, 2021, pp. 115--131.

\bibitem{goel2021symmetry}
A.~Goel and K.~Sakallah, ``On symmetry and quantification: A new approach to verify distributed protocols,'' in \emph{NASA Formal Methods Symposium}.\hskip 1em plus 0.5em minus 0.4em\relax Springer, 2021, pp. 131--150.

\bibitem{yao2021distai}
J.~Yao, R.~Tao, R.~Gu, J.~Nieh, S.~Jana, and G.~Ryan, ``$\{$DistAI$\}$:$\{$Data-Driven$\}$ automated invariant learning for distributed protocols,'' in \emph{15th USENIX symposium on operating systems design and implementation (OSDI 21)}, 2021, pp. 405--421.

\bibitem{yao2022duoai}
J.~Yao, R.~Tao, R.~Gu, and J.~Nieh, ``$\{$DuoAI$\}$: Fast, automated inference of inductive invariants for verifying distributed protocols,'' in \emph{16th USENIX Symposium on Operating Systems Design and Implementation (OSDI 22)}, 2022, pp. 485--501.

\bibitem{xia2025discovering}
Y.~Xia, D.~Sur, A.~S. Pingle, J.~V. Deshmukh, M.~Raghothaman, and S.~Ravi, ``Discovering likely invariants for distributed systems through runtime monitoring and learning,'' in \emph{International Conference on Verification, Model Checking, and Abstract Interpretation}.\hskip 1em plus 0.5em minus 0.4em\relax Springer, 2025, pp. 3--25.

\bibitem{cao2018cmp}
J.~Cao, Y.~Li, and J.~Pang, ``L-cmp: An automatic learning-based parameterized verification tool,'' in \emph{Proceedings of the 33rd ACM/IEEE International Conference on Automated Software Engineering}, 2018, pp. 892--895.

\bibitem{emerson2003exact}
E.~A. Emerson and V.~Kahlon, ``Exact and efficient verification of parameterized cache coherence protocols,'' in \emph{Advanced Research Working Conference on Correct Hardware Design and Verification Methods}.\hskip 1em plus 0.5em minus 0.4em\relax Springer, 2003, pp. 247--262.

\bibitem{hawblitzel2015ironfleet}
C.~Hawblitzel, J.~Howell, M.~Kapritsos, J.~R. Lorch, B.~Parno, M.~L. Roberts, S.~Setty, and B.~Zill, ``Ironfleet: proving practical distributed systems correct,'' in \emph{Proceedings of the 25th Symposium on Operating Systems Principles}, 2015, pp. 1--17.

\bibitem{dill1996mur}
D.~L. Dill, ``The mur $\phi$ verification system,'' in \emph{Computer Aided Verification: 8th International Conference, CAV'96 New Brunswick, NJ, USA, July 31--August 3, 1996 Proceedings 8}.\hskip 1em plus 0.5em minus 0.4em\relax Springer, 1996, pp. 390--393.

\bibitem{cimatti2002nusmv}
A.~Cimatti, E.~Clarke, E.~Giunchiglia, F.~Giunchiglia, M.~Pistore, M.~Roveri, R.~Sebastiani, and A.~Tacchella, ``Nusmv 2: An opensource tool for symbolic model checking,'' in \emph{Computer Aided Verification: 14th International Conference, CAV 2002 Copenhagen, Denmark, July 27--31, 2002 Proceedings 14}.\hskip 1em plus 0.5em minus 0.4em\relax Springer, 2002, pp. 359--364.

\end{thebibliography}

\end{document}